\documentclass[11pt]{article}

 \usepackage{amsmath}
 \usepackage{graphicx}
 \usepackage{amssymb}
 \usepackage{amsfonts}
 \usepackage{latexsym}

\renewcommand{\baselinestretch}{1.2}
\def\beq{\begin{eqnarray}}
\def\eeq{\end{eqnarray}}
\def\ln{\,\mbox{ln}\,}

\def\al{\alpha}
\def\be{\beta}

\def\de{\delta}

\def\si{\sigma}

\def\La{\Lambda}

\begin{document}

\begin{center}

{\large\sc On the scaling rules for the anomaly-induced effective action
of metric and electromagnetic field}
\vskip 4mm


\textbf{Ana M. Pelinson}$^{\,a,}
$\footnote{
E-mail: ana.pelinson@gmail.com}
\quad
\textbf{and}
\quad
\textbf{Ilya L. Shapiro}$^{\,b,}
$\footnote{
E-mail: shapiro@fisica.ufjf.br. On leave from Tomsk State
Pedagogical University, Tomsk, Russia.}
$\,,\,\,$
\vskip 6mm
$^{\,a}$ Departamento de F\'{\i}sica, CFM,
Universidade Federal de Santa Catarina, 88040-900,
Florian\'{o}polis, SC, Brasil
\vskip 4mm
$^{\,b}$ Departamento de F\'{\i}sica, ICE,
Universidade Federal de Juiz de Fora, 36036-330, MG, Brazil
\end{center}

\vskip 12mm

\begin{quotation}
\noindent {\large {\it Abstract}}.
\quad
The anomaly-induced effective action is a useful tool for
deriving the contributions coming from quantum effects of
massless conformal fields. It is well-known that such
corrections in the higher derivative vacuum sector of the
gravitational action provide the same exponential inflation
(Starobinsky model) as the cosmological constant term. At
the same time, the presence of a classical electromagnetic
field breaks down the exponential solution. In this paper
we explore the role of the anomaly-induced term in the
radiation sector and, furthermore, derive the ``equation of
state'' and the scaling laws for all terms in the Einstein
equations. As one could expect, the scaling law for the
vacuum  anomaly-induced effective action is the same as
for the cosmological constant.
\end{quotation}
\vskip 4mm

\section{Introduction}

It is well-known that the conformal anomaly is useful for
various applications Quantum Field Theory. In particular,
the anomaly-induced effective action has been explored in
the cosmological setting about three decades ago \cite{fhh}.
Soon it was discovered
that, in the absence of matter fields or radiation, the
quantum anomaly-induced contributions lead to the Starobinsky
model of inflation \cite{star}
(see also \cite{mm,star1,ander,vile,anju,hhr} for an
alternative work and further developments).
The traditional version of this inflationary model is based on
the unstable exponential solution \cite{star}, that implies
some special choice of the number of the quantum fields with
different spins \cite{anju}. For instance, the present-day
Universe with (presumably) only photon being an active quantum
field, or the early universe where the active quantum content
is described by the Minimal Standard Model of particle physics,
satisfy the condition of unstable inflation.

An alternative possibility is to
consider the supersymmetric matter content of active fields,
that leads to the stable inflationary solution at the initial
stage of inflation. The transition from stable to unstable
inflationary regimes can be associated to the decoupling of
the massive $s$-particles
\cite{insusy,shocom} in the universe where the
inflation is slowing down because of the quantum effects of
massive fields \cite{shocom}. The same effect holds in the
presence of the cosmological constant, which actually plays
only a small role in this story \cite{asta,asta-s,ana-so}.
In both cases
of stable and unstable inflationary solutions one usually
assume that the universe is empty, that means there are
no matter fields and/or radiation.

All the mentioned massive or
massless fields of different spins are virtual ones, they
manifest themselves only through their contributions to the
vacuum action. If the real radiation is present, there is
no exponential solution for the conformal factor $a(t)$ and
the last tends to the corresponding FRW solution if the
particle content corresponds to the unstable case and if
the initial data are chosen in an appropriate way \cite{fhh}.
After a while, the dynamical system describing the universe
enters the regime where the effect of higher derivative terms
becomes negligible \cite{vile,anju,asta}, and the behaviour
of the conformal factor is essentially the same as in the
purely classical universe dominated by classical radiation
content.

The above conclusion is based on the analysis of the theory with
the action which includes Einstein-Hilbert term, classical
radiation and higher-derivative anomaly-induced gravitational
contributions. However, there may be one missed component in
this consideration.
In the case when the background radiation is present, one may
need to take into account also the anomaly-induced contribution
to the electromagnetic part. Indeed, the classical radiation
does decouple from the equation for the conformal factor, and
the radiation density manifests itself only via the first
integral of this equation, that is the first Friedmann equation.
So, it looks interesting to check what is the effect of the
anomaly-induced term, e.g., whether it is capable to produce
some significant change in the acceleration of the universe.
The same problem can be explored, also, for the
radiation-dominated epoch afterthe inflation ends.

One can consider a bit more general formulation of the
problem. We can derive the ``equation of state'' for all
components of the gravitational action, namely the
Einstein-Hilbert term, cosmological constant term,
free radiation, quantum anomaly-induced contribution to
the radiation part and quantum anomaly-induced contribution
to the vacuum part. Then we can check how the corresponding
``energy densities'' depend on the scale factor. For instance,
the comparison of these dependencies for the anomaly-induced
vacuum terms and the cosmological constant can better explain
why the two kind of vacuum actions produce similar exponential
behaviour. This issue may be also interesting in view of
the recent attempts to deal with the cosmological constant
problem by taking the anomaly-induced contributions into
account \cite{fossil,TUZh}.

The paper is organized as follows. In Sect. 2 we write
down the anomaly-induced terms in both gravitational and
electromagnetic sectors and consider the relations between
timelike and spacelike components of the diagonalized
equations for the metric (which are Energy-Momentum Tensors
in the electromagnetic field case). These relations can be
seen as equation of state for all the terms in the modified
Einstein equations in the cosmological setting. In  Sect. 3
we explore what is the effect of the anomaly-induced
electromagnetic term for the rate of expansion of the
universe in the two different situations, namely when
the higher derivative metric dependent terms are present
or not. Finally, in  Sect. 4 we draw our conclusions and
discuss some possible applications of the results.

\section{Classical and anomalous terms in the effective action}

The conformal anomaly is the typical theoretical phenomenon
for massless conformal invariant quantum
fields on some nontrivial external background. In case of
massless conformal fields the action of vacuum (gravitational
one) has to include, at least, the conformal invariant higher
derivative part (see, e.g., \cite{book} for the introduction
and \cite{Poimpo} for a recent review of Quantum Field Theory
in curved space)
\beq
S_{HD}\,
= \,\int d^4x\sqrt{-g}\,\big(
a_1C^2 + a_2E + a_3{\nabla^2} R \big)\,.
\label{S HD}
\eeq
Here $\,C^2=R_{\mu\nu\al\be}^2 - 2 R_{\al\be}^2 + (1/3)\,R^2\,$
is the square of the Weyl tensor and
$\,E = R_{\mu\nu\al\be}^2 - 4 R_{\al\be}^2 + R^2\,$
is the integrand of the Gauss-Bonnet topological term. The terms
in the Lagrangian of (\ref{S HD}) satisfy the conformal Noether
identity and, furthermore, do not affect the dynamical equation
for the conformal factor of the metric. At the quantum level,
however, the conformal symmetry is violated and this also
affects the cosmological solution.

In the cosmological setting, massless conformal invariant
quantum fields corresponds to the early epoch
when the energy of the photons is much greater than masses
of at least some of the charged spinor fields. This
condition can be easily satisfied in the inflationary
period, especially in the framework of the Starobinsky model,
which has, usually, very high values of the typical energies
at the end of inflationary period. Furthermore, this condition
can be fulfilled in the radiation-dominated period after
inflation, where many massive fields approximately can be
approximately treated as massless.

\subsection{Anomaly-induced terms}

Consider the approximation of massless fields. In case of
both gravitational and electromagnetic background fields,
the conformal anomaly has the form
\beq
<T_\mu^\mu> \,=\, - \,\big(wC^2 + bE + c{\nabla^2} R+\be F^2\big)\,\,,
\label{anomaly}
\eeq
where $\,F^2=F^2_{\mu\nu}\,$ is square of the strength
tensor of the electromagnetic fields, $\,w,\,b,\,c\,$ are
the $\,\beta$-functions for the parameters of the vacuum
action and $\be$ is proportional to the electromagnetic
charge $\,\beta$-function. At one loop order, using the
Minimal Subtraction scheme of renormalization, we get
\beq
\be\,=\,- \frac{2e^2}{3(4\pi)^2}\sum_f N_f
\,-\, \frac{e^2}{6(4\pi)^2}\sum_s N_s
\label{beta}
\eeq
as a sum over charged fermions and scalars with the
multiplicities $N_f$ and $N_s$
correspondingly.~\footnote{The detailed discussion of
the anomaly-induced action of electromagnetic field and
its relation to the more general result coming from the
physical renormalization scheme can be found in
\cite{FormQED}, see also \cite{Mottola-08,Coriano}.}
The one-loop values of $\,w,b\,$ and $\,c\,$ can be
found, e.g. in \cite{birdav,book,Poimpo}.

It is well known that taking into account the conformal
anomaly in the cosmological case leads to the Starobinsky
exponential solution \cite{star} for the conformal factor,
if there are no matter fields. At the same time, if the
radiation is present, there is no such solution. One
can naturally ask whether the anomalous electromagnetic
term in (\ref{anomaly}) can change this situation. And
more general, whether this term can affect the expansion
of the universe at the early stage of its history.

In order to use field quantities in the cosmological
setting, one has to perform some space averaging.
Obviously, $\,\langle F^2 \rangle \sim \langle E^2 \rangle
- \langle H^2\rangle\,$
equals zero for a free radiation. But this does not apply,
e.g., to the
radiation-dominated early universe, because in this case
there is also a hot plasma of other particles and the
content of the universe does not reduce to a free
electromagnetic radiation.
As a qualitative simplest estimate we shall suppose that
$F^2 \neq 0$ and set its scale-factor dependence in
accordance to its conformal property. One can assume,
for instance, that at some fixed scale the magnitude of
this term is proportional to the $\rho^0_r$, that is the
classical radiation energy density. This radiation density
is supposed to describe not only electromagnetic fields,
but also a hot plasma which fills the Universe. It is
important to note that such non-trivial material content
of the universe is indeed possible at the last stage of
the stable inflation, where we observe oscillations of
the conformal factor \cite{shocom} and, consequently,
production of photons and charged particles.

The anomaly-induced effective action can be easily
derived as a functional of the new variables
$\,{\bar g}_{\mu\nu}\,$ and $\,\sigma$, where
$\,{g}_{\mu\nu} = {\bar g}_{\mu\nu}\cdot e^{2\sigma}$
and the metric $\,{\bar g}_{\mu\nu}\,$ has fixed
determinant. Disregarding the conformal
invariant term in the effective action we arrive at
the following expression \cite{rei}:
\begin{eqnarray}
{\bar \Gamma} = \int d^4 x
\sqrt{-{\bar g}} \,\{w\sigma{\bar C}^2 + b\sigma
({\bar E}-\frac23 {\bar \nabla}^2 {\bar R})
+ 2 b\,\sigma{\bar \Delta}\sigma +
\,\be\sigma{\bar F}^2\}\,
- \,\frac{3c+2b}{36}\,\int d^4 x\sqrt{-g}\,R^2\,,
\label{quantum}
\end{eqnarray}
where
$\,{\bar F}^2 = {\bar g}^{\mu\al} {\bar g}^{\nu\be}
F_{\al\be}F_{\al\be}=e^{-4\si}F^2$.
The expression (\ref{quantum}) is the quantum correction
to the classical action of vacuum. Let us note that the
covariant (nonlocal and local) forms of the anomaly-induced
action are well-known \cite{rei,FormQED,Mottola-08}, but
the eq. (\ref{quantum}) is sufficient for our present
purposes. The total action has the form
\beq
S_t \,=\,\,S_{EH}+{\bar \Gamma}+{\rm conf.\, invariant\, terms}\,,
\label{totalaction}
\eeq
where the conformal invariant terms include the classical
actions of radiation and of hot charged particles and
$S_{EH}$ is the Einstein-Hilbert term
\beq
S_{EH}\,
= \,-\,\frac{1}{16\pi G}\,\int d^4x\sqrt{-g}\,(R + 2\La)\,.
\label{Einstein}
\eeq

In the case of the cosmological, FRW metric, the classical
massless fields decouple from gravity. Taking into account
the conformal properties, it proves useful to rewrite the
expression (\ref{totalaction}) in a more detailed form
\beq
S_t \,=\,\,S_{EH}
+ S_{HD}
+ S^0_{r}
+ {\bar \Gamma}_{HD}
+ {\bar \Gamma}_{\beta}\,.
\label{tact}
\eeq
Here $S_{HD}$ and $S^0_r$ are
classical higher derivative metric and radiation (including
massless charged fields) conformal invariant actions.
${\bar \Gamma}_{HD}$ and ${\bar \Gamma}_{\beta}$ are parts of
anomalous action (\ref{quantum}). In what follows we will use
the same indications for all quantities, including the trace
of the stress-tensor,
\ $T_i=(T_{EH},\,T_{HD},\,{\bar T}_{HD},\,T^0_r,\,T_\beta)$, \
energy density \ $\rho_i$ \ and pressure \ $p_i$, with
\ $T_i=\rho_i-3p_i$ \ in the corresponding reference frame. On the
top of that, we will sometimes use notation for the total
expression in the radiation sector, like
\ $\rho_r=\rho^o_r+\rho_\beta$.

\subsection{Energy density and pressure}

Consider the stress-energy tensor, whose components
are given by the variational derivative of the total
effective action (\ref{tact}),
\beq
T^{\alpha \beta}=-\frac{2}{\sqrt{-g}}\frac
{\de S_t}{\de g_{\alpha\beta}}\,.
\label{T equation}
\eeq
In order to calculate separately the contributions of the
Einstein-Hilbert term from the terms with high derivatives (HD)
which come from the quantum contributions and the electromagnetic
one, we present the trace of the stress-energy tensor in the form
\beq
T =
\frac{1}{a^3}\frac{\de S_t}{\de a}
\,=\, T_{EH} + T^0_{r} + T_{HD} + {\bar T}_{HD} + T_{\beta}
\,=\, \rho_t-3p_t\,,
\label{T components}
\eeq
where
\beq
\rho_t=\rho_{EH} + \rho^0_{r}  + \rho_{HD} + \rho_{HD}+ \rho_{\beta}
\,=\,-\,T_0^0
\label{rho components}
\eeq
and (in Cartesian coordinates)
\beq
p_t\,=\,p_{EH} + p^0_{r}+ p_{HD} + {\bar p}_{HD} + p_{\beta} =T_1^1 \,=\, T_2^2=T_3^3\,
\label{p components}
\eeq
are density-like and pressure-like components, correspondingly.
The indices and bars of all quantities are in accordance with the
ones of the actions in the {\it r.h.s.} of (\ref{tact}).
We find useful to introduce such notations even for the
Einstein-Hilbert term, despite the physical sense of the
quantities is different in this case (as it is, of course,
for the higher derivative vacuum terms, too).

Our purpose is to see how all $\rho$'s and  $p$'s depend on the
scale factor $a(t)$ and also how they behave during and after
the inflationary period. One can find the densities for the
components by assuming that the conservation law is satisfied
separately for each of the stress-energy tensors
\ $T_i = (T_{EH}, T^0_r, T_{HD}, {\bar T}_{HD}, T_\be)$ \
in (\ref{T components}). In terms of the
cosmic scale factor the conservation law can be expressed as
\beq
d\,(\rho_i \,a^3) =-p_i \,d(a^3)\,,
\quad \mbox{where} \quad
p_i=\frac{\rho_i-T_{i}}{3}\,.
\label{conslaw}
\eeq

Following this standard procedure, we can immediately see that
since the trace of the classical radiation stress tensor is zero,
the equation of state is as it is supposed to be,
\beq
T^0_{r}=3p^0_r-\rho^0_r=0\,, \quad \mbox{hence} \quad
p^0_r=\frac{\rho^0_r}{3}\,.
\label{radEOS}
\eeq
Correspondingly, this term does not contribute to the
equation of motion for $a(t)$.

The first observation is that, since the trace is zero for
the $S_{HD}$ term, the corresponding equation of state is
exactly the same as for the free radiation,
$\,p_{HD}=\rho_{HD}/3$. For other three terms we obtain, in
terms of conformal time \ $\eta$ \ (as usual, $dt=a(\eta)d\eta$),
\beq
T_{EH}=\frac{3}{4\pi G} \left[\frac{a^{\prime\prime}}{a^3}
-\frac{2 \Lambda}{3}\right]\,,
\eeq
\beq
{\bar T}_{HD}
\,=\, 6c \left[-\frac{a^{\prime\prime\prime\prime}}{a^5}
+ 4\frac{a^{\prime\prime\prime}a^{\prime}}{a^6}
+ 3\left(\frac{a^{\prime\prime}}{a^3}\right)^2
- 6\frac{a^{\prime\prime} a^{\prime\, 2}}{a^7}\right]
- 24b\left[ \left(\frac{a^{\prime}}{a^2}\right)^4
- \frac{a^{\prime\prime} a^{\prime \, 2}}{a^7} \right]\,,
\eeq
and
\beq
T_{\beta}= \frac{\beta {\bar F^2} }{a^4}\,.
\eeq

Using Eqs. (\ref{conslaw}), the solution for
all the densities $\rho_i$ come from the differential equations
of the form
\beq
\frac{d\rho_i}{da^3}+\frac{4}{3}\frac{\rho_i}{a^3}=\frac{T_i}{3a^3}\,.
\label{density equation}
\eeq
A general solution for this non-homogeneous equation
(\ref{density equation}) is
\beq
\rho_i(a)=C(a)a^{-4}\,,
\label{density solution}
\eeq
where the coefficient $C(a)$ is obtained by the integration of
\beq
\frac{d C}{d\eta} = T_i \,a^3\,a^{\prime}\,.
\label{coef. equation}
\eeq

Integrating (\ref{coef. equation}) for each of remaining
stress-energy tensor components above and substituting them into
Eq. (\ref{density solution}), we arrive at the following results:
\beq
\rho_{EH} &=&
\frac{3}{8\pi G}\,\left(\frac{a^{\prime}\,^2}{a^4}
- \frac{\Lambda}{3}
\right)\,,
\nonumber
\\
p_{EH} &=& - \frac{1}{8\pi G}\,
\left(2\,\frac{a^{\prime\prime}}{a^3}
-\frac{a^{\prime}\,^2}{a^4}-\Lambda\right)\,,
\label{EOS EH}
\eeq
\beq
{\bar \rho}_{HD}
&=& - 6c\left[\frac{a^{\prime\prime\prime}a^{\prime}}{a^6}
-\frac{1}{2}\left(\frac{a^{\prime\prime}}{a^3}\right)^2
-2\frac{a^{\prime\prime}a^{\prime\, 2}}{a^7}
\right]+6b \left(\frac{a^{\prime}}{a^2}\right)^4\,,
\nonumber
\\
{\bar p}_{HD} &=&
-2c\left[5\frac{a^{\prime\prime\prime}a^{\prime}}{a^6}
- \frac{a^{\prime\prime\prime\prime}}{a^5}
+\frac{5}{2}\left(\frac{a^{\prime\prime}}{a^3}\right)^2
-8\frac{a^{\prime\prime}a^{\prime\, 2}}{a^7}\right]
+8b \left[3\left(\frac{a^{\prime}}{a^2}\right)^4
-\frac{a^{\prime\prime}a^{\prime\, 2}}{a^7}\right]\,,
\label{EOS HD}
\eeq
\beq
\rho_{\beta}=\frac{\beta\,{\bar F}^2}{a^{4}}\ln a\,,
\qquad
p_{\beta}=\frac{\beta\,{\bar F}^2}{3\,a^{4}}(\ln a - 1)\,.
\label{EOS R}
\eeq
Indeed, the equations (\ref{EOS HD}) derived here are well-known,
they are exactly the same as the ones obtained in \cite{fhh}, and
also recalculated in \cite{hhr}. It is easy to see that these
formulas are quite different from the ones for the cosmological
constant in (\ref{EOS EH}). One can expect that for the general
form of $a(\eta)$ these two different equations of state will
definitely produce different contributions. However, in short
we will see the effect of the two terms is equal for the
exponential inflation case.

\section{Cosmological solutions with anomalous terms}

Here we consider the effect of radiation anomalous term on the
behaviour of the conformal factor of the metric, in the
framework of a FLRW cosmology, with and without higher
derivative anomalous terms.

\subsection{Stable anomaly-induced inflation with radiation term}

As a first step, consider the equation of motion including the
higher derivative terms. The most useful choice of variable is
the conformal factor as a function of cosmic time, $\sigma(t)$.
The last quantity is defined as \ $\sigma = \ln a$.
The equation of motion can be obtained from the $00$-component
\cite{fhh,star} or directly from the trace $T=\rho_t-3p_t=0$
\cite{anju}. In terms of $\tau=t/t_{Pl}=M_{Pl}t$, where $t_{Pl}$
is the Planck time unit $t_{\rm Pl}\simeq 5.3\times 10^{-44} s$,
the equation has the form
\beq
\stackrel{....}{\si}
&+& 7 \stackrel{.}{\si}\stackrel{...}{\si}
\,+ \,4 \Big(3 - \frac{b}{c} \Big)
    \,\stackrel{.}{\si}^2\stackrel{..}{\si}
\,+\, 4 \stackrel{..}{\si}^2
\,- \,\frac{4b}{c}\,\stackrel{.}{\si}^4
\nonumber
\\
&-& \frac{1}{ c}\,\Big(\stackrel{..}{\si} \,+ \, 2\stackrel{.}{\si}^2-\frac{2}{3}\frac{\rho_{\Lambda}}{M_{Pl}^4}\Big)
-\,\frac{1}{6c}\left(\frac{\be {\bar F}^2}{{M_{Pl}^4}}\right)e^{-4\si}
\,=\,0\,.
\label{great}
\eeq
In this equation the contribution of the cosmological constant
term is written in terms of vacuum energy density
$\rho_{\rm \Lambda}= \Lambda/(8\pi G)= \Lambda M_{Pl}^2$, where
$M_{Pl}=1/\sqrt{8\pi G}= 2.44 \times 10^{18} \, {\rm GeV}$ is the
reduced Planck mass.

The direct inspection shows that, in the presence of the
$\,\be {\bar F}^2$-term, there is no exponential solution.
This demonstrates that the importance of vacuum for such
solution hold also when we take the anomaly in the radiation
sector into account.

The next question is what is the role of the $\be {\bar F}^2$-term
for the case of a stable inflation. In principle, one can expect
two different situations: \ (i) The anomalous term slows down
the exponential inflation, as it happens with the terms
generated by the quantum effects of massive light fields
\cite{shocom,asta}; \ (ii) The anomalous $\be {\bar F}^2$-term
decreases very fast and soon becomes negligible. The numerical
analysis show that this last behaviour actually takes place.
For the illustration we present the corresponding plots for the
case of Minimal Supersymmetric Standard Model (MSSM) in Fig. 1.
As we have already mentioned in the Introduction, the
supersymmetric particle content is the most interesting here,
because it provides stable inflation, making the possible effect
of the radiation term (or the absence of such effect) the most
explicit and clear.

In order to find the cosmological evolution of all the densities,
we solve eq. (\ref{great}) numerically and then replace the
solution into the expressions which directly follow, in particular,
from \ (\ref{EOS EH}), (\ref{EOS HD}) and (\ref{EOS R}).
We present these results, in Fig. 1, as functions of $\tau=t/t_{Pl}$,
where
\beq
\frac{\rho_{EH}(\tau)}{M_{Pl}^4}=3\stackrel{.}{\si}\,^2
-\frac{\rho_{\Lambda}}{M_{Pl}^4}\,,
\eeq
\beq
\rho^0_r(\tau)=\,e^{-4\sigma}\,{\rho_r^0}(\tau=0)\,,
\label{rho_R}
\eeq
\beq
\frac{{\bar \rho}_{HD}(\tau)}{M_{Pl}^4}
= - 6c\left(\stackrel{.}{\si}\stackrel{...}{\si}
+ 3\,\stackrel{.}{\si}^2\stackrel{..}{\si}
- \frac{1}{2}\stackrel{..}{\si}^2
- \frac{b}{c}\,\stackrel{.}{\si}^4 \right) \,,
\eeq
\beq
\frac{\rho_{\beta}(\tau)}{M_{Pl}^4}
\,=\,\frac{\beta\,{\bar F}^2}{M_{Pl}^4}\,\sigma \,e^{-4\sigma}\,,
\eeq
for the MSSM particle content with $N_{1,1/2,0}=(12,48,104)$.

\begin{figure}
  \centering
{\includegraphics[width=80mm]{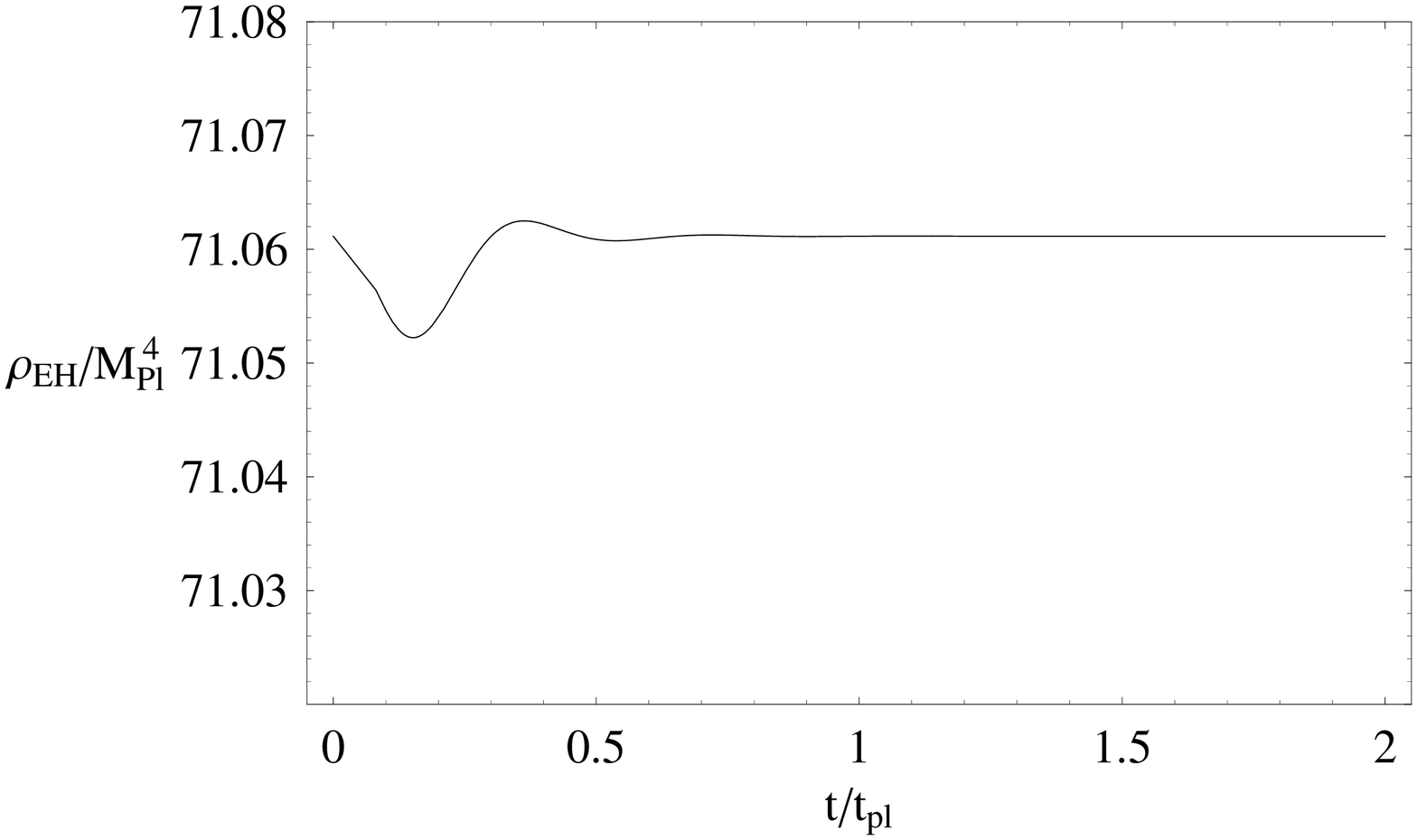}}
{\includegraphics[width=80mm]{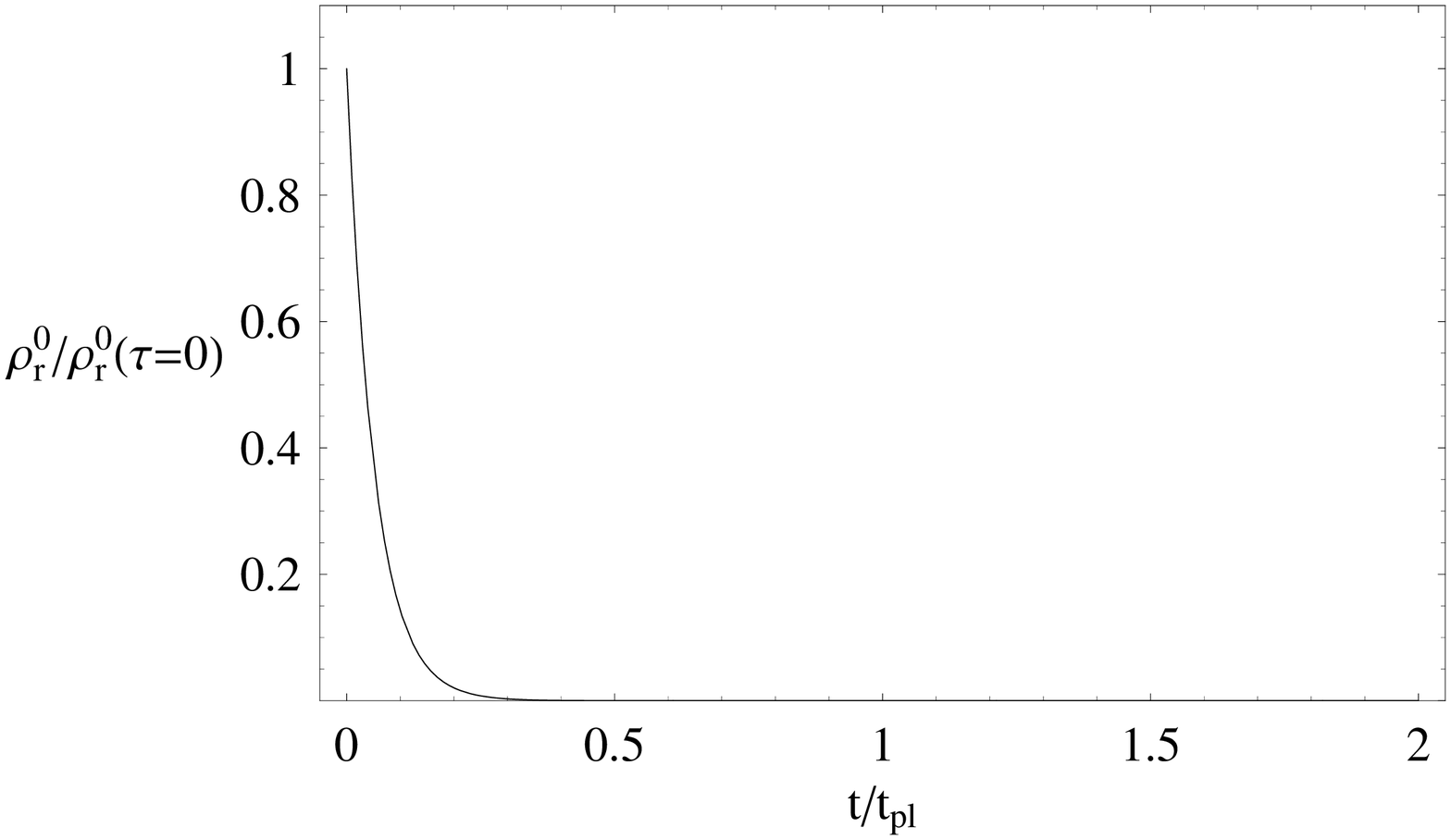}}
{\includegraphics[width=80mm]{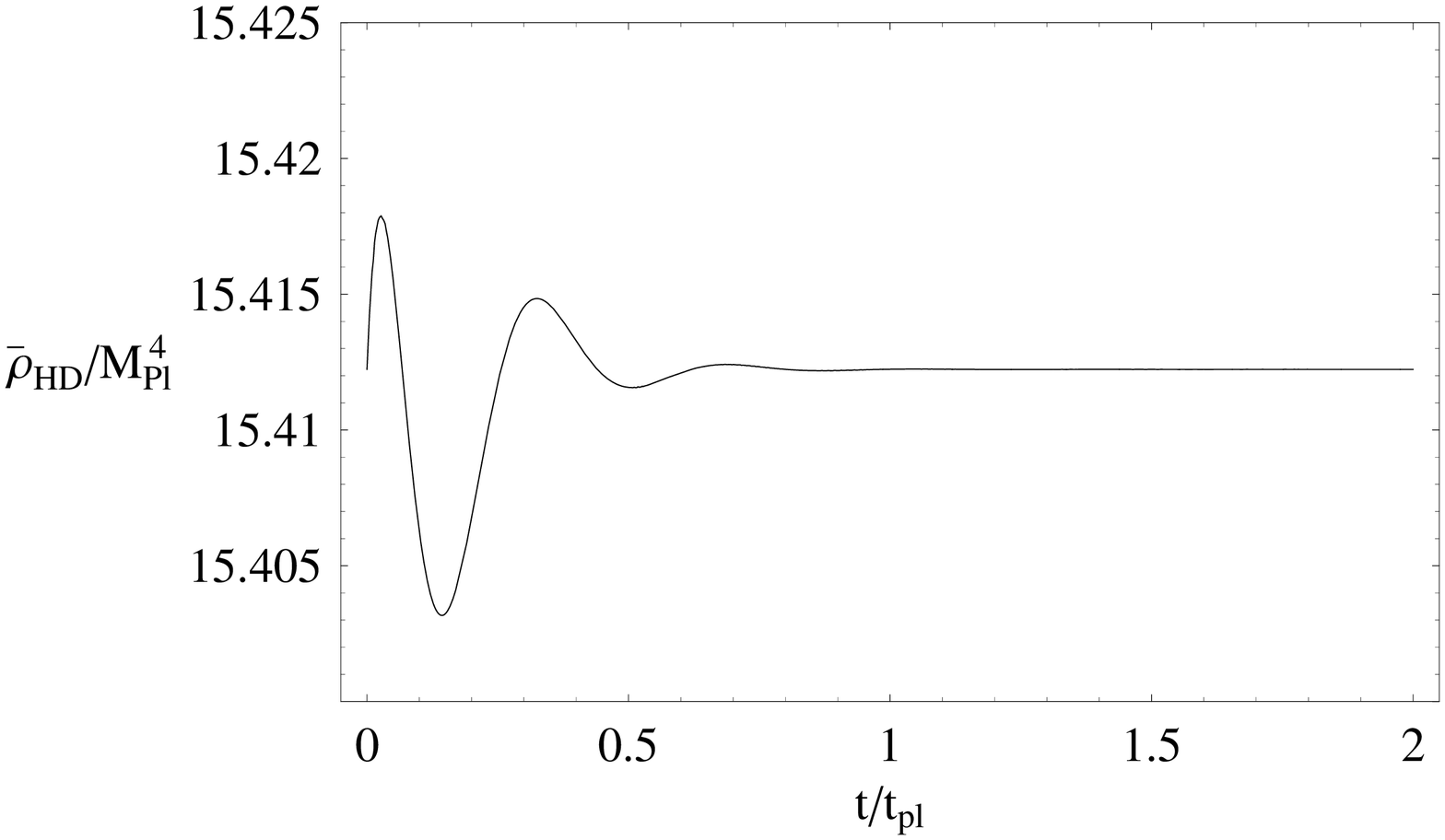}}
{\includegraphics[width=80mm]{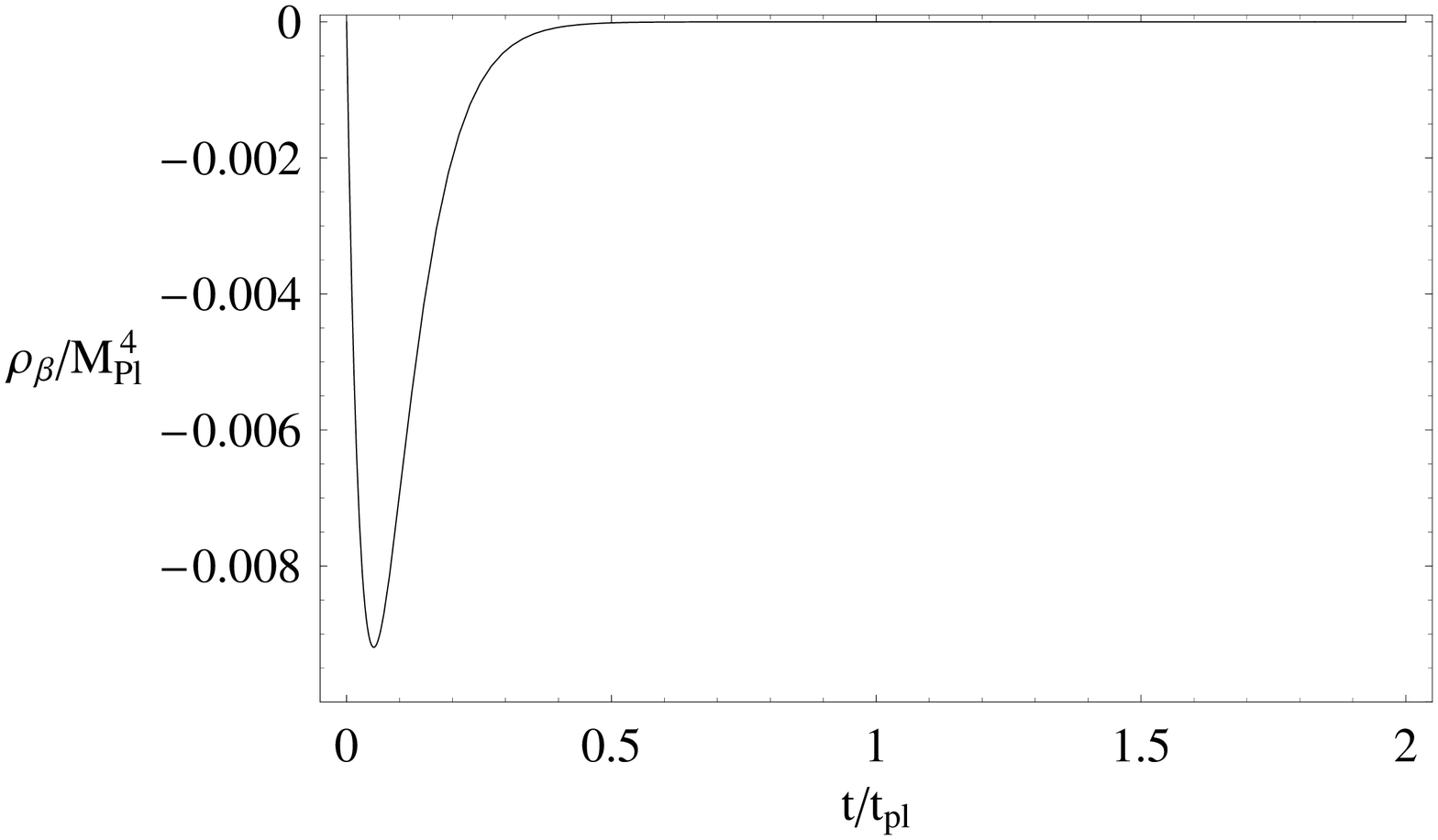}}
\caption{We have assumed here the MSSM particle content with
$N_{1,1/2,0}=(12,48,104)$ and took the numerical value
$\,\,\beta {\bar F}^2 = - 0.1$. The plots for ${\bar \rho}_{HD}$
and $\rho_{EH}$ rapidly tend to constants, because the
inflation is stable and in the exponential regime the
curvature components behave like constants. In
this regime, it shows exactly the same scaling law as the
density of the cosmological constant. }
\end{figure}

\subsection{Radiation-dominated evolution after inflation}

In this section we shall investigate the effect of
quantum corrections at the period after the anomaly-induced
inflation ends and the higher derivative terms in (\ref{quantum})
become negligible. Then the relevant part of the total action
has the form
\beq
S_t \,=\,-\,\frac{1}{16\pi G}\,\int d^4x\sqrt{-g}\,R
+ S_{class.matter}
+ \int d^4 x \sqrt{-{\bar g}}\,\be\,\sigma{\bar F}^2\,,
\label{total}
\eeq
where $S_{class.matter}$ is the classical action of
the matter fields. We are interested in the period when
matter and radiation are very hot and can be treated as
conformal. Then, the classical massless fields in
$\,S_{class.matter}\,$ decouple from the conformal
factor of the metric and the effective equation
of motion, in terms of conformal time, has the form
\beq
\frac{3}{4\pi G}\,a^{\prime\prime}
+ \frac{\be{\bar F}^2}{a}\,=\,0\,.
\label{equation1}
\eeq

Let us, as before, denote the
derivative with respect to the cosmic time $t$ by a point.
Then Eq. (\ref{equation1}) becomes
\beq
{\ddot a} + \frac{{\dot a}^2}{a}
- \frac{\mu^2}{2a^3} = 0\,.
\label{equation}
\eeq
In the last equation we have introduced a useful
notation\footnote{Let us note that
${\bar F}^2=2(\langle H^2\rangle -\langle E^2\rangle)$
is typically positive and
that, according to our definition (\ref{beta}), $\be <0$.}
\beq
\mu^2 = - \frac{8\pi G\be{\bar F}^2}{3}\,.
\label{newnotation}
\eeq
As a first step in solving Eq. (\ref{equation}) we
find the relation between $a(t)$ and $H = {\dot a}/a$
\beq
H^2 \,=\,\frac{C+\mu^2\,\ln a}{a^4}\,,
\label{Hubble}
\eeq
where $C$ is an integration constant. In order to clarify
the sense of this constant, let us consider the standard
classical model with $\mu=0$. In this case, from Eq.
(\ref{Hubble}) follows $a(t)=C(t-t_0)^{1/2}$, where
$t_0$ is some fixed instant of time. Different choices
of  $t_0$ can be compensated by the renormalization
of $C$, so we set $t_0=0$. On the other hand, by solving 
the Friedmann equation we obtain
\beq
a(t) \,=\,\Big[\frac{32\pi G\rho^0_r(t=0)}{3}\Big]^{1/4}\cdot
a_0\cdot \sqrt{t}\,,
\label{energy}
\eeq
where $\rho^0_r(t=0)$ and $a_0$ are the energy density
of the electromagnetic field and the scale factor of the
metric at the instant $t=0$. The comparison of the two
expressions for $a(t)$ lead to the relation
$\,C=\frac{8\pi G\rho_r^0(t=0)}{3}$. It is natural to fix
that the $a_0$ corresponds to the fiducial metric
${\bar g}_{\mu\nu}$. In what follows we put $a_0=1$.
Then the elements of the solution (\ref{Hubble}) satisfy
the relation $\mu^2 \ll C$ if the anomalous contribution
and energy density satisfy the relation
$|\be {\bar F}^2| \ll 4\rho_r^{(0)}$. As we have
indicated above, this relation is quite natural,
because for the free radiation ${\bar F}^2=0$ and
the presence of the anomalous term is due to the
interaction with other fields which have energy
density much smaller than the one of radiation.

Finally, the general analytic solution of (\ref{equation})
can be presented in the form
\beq
t \,=\, \frac{2\,e^{-2C/\mu^2}}{\mu}\,
\int\limits_{\sqrt{C}/\mu}^{\sqrt{\si + C/\mu^2}}e^{2z^2}\,dz
\,=\,\frac{\,e^{-2C/\mu^2}}{\mu}\sqrt{\frac{\pi}{2}}\,\Big[
{\rm Erfi} \big(\sqrt{2\si + {2C}/{\mu^2}}\big)
\,-\, {\rm Erfi}\big({\sqrt{2C}}/{\mu}\big) \Big]\,.
\label{general solution}
\eeq
The disadvantage of this formula is that it becomes
singular in the classical limit $\mu \to 0$. In order
to solve this difficulty, one can derive an aproximate
solution by
treating the term with $\mu$ as a small perturbation,
\beq
\sqrt{C}\,t\,\cong\,\frac{a^2}{2}\,
\Big(1- \frac{\mu^2}{2C}\cdot \ln a\Big)\,.
\label{solution}
\eeq
As one can see from the last relation, the expansion of the
universe performs slightly faster as a result of the quantum
effects related to the electromagnetic anomalous term $\be F^2$.

When the universe expands, the radiation temperature is
decreasing. It is instructive to find the temperature relations
when the quantum term in the solution (\ref{solution})
is relevant. The lower bound for the relevant temperature
is defined by the energy corresponding
to the moment when the lightest charged fermion decouples and the
upper bound is the energy scale when the inflation ends and the
higher derivative terms in Eq. (\ref{quantum}) become
negligible. In the framework of the modified Starobinsky model
\cite{insusy,shocom,asta}, the scale of the graceful
exit from the anomaly-induced inflation depends on the
scale of the supersymmetry breaking, and may vary from
$H=300\,GeV$ to $H=10^{14}\,GeV$ for different gauge
theories. Let us notice that, contrary
to the case of a purely gravitational background, the
mixed electromagnetic-gravitational background corresponds
to the loop diagrams with external lines of both
electromagnetic potential and metric perturbations.
Indeed, the decoupling scale is defined by the energy
of the photons which is much greater than the energy of
the gravitons.

In order to find the temperature of the radiation
corresponding to (\ref{solution}), we can use the
Friedmann equations
\beq
\frac{{\dot a}^2}{a^2} \,=\,H^2\,=\,\frac{8 \pi G}{3}\,\rho_r
\,,\qquad
\frac{{2\ddot a}}{a}\,+\,\frac{{\dot a}^2}{a^2}
\,=\,-\,8 \pi G\,p_r\,,
\label{Friedmann}
\eeq
where $p_r$ is the radiation pressure
and the thermodynamical relation is close to
$\,\rho_r \approx \rho^0_r=\frac{\pi^2}{15}\,T^4$.
Then we arrive at the formula
\beq
T^4 \,=\,\frac{45}{ \pi^2}\,H^2\cdot M_{Pl}^2\,.
\label{T}
\eeq
The next requires the equation of state for the anomalous
term, which was obtained in the previous section. Using
this result directly leads us to
\beq
\rho_r\,=\,\frac{3}{8 \pi G}\cdot\frac{C+\mu^2\ln a}{a^4}
\label{density}
\eeq
and
\begin{eqnarray}
p_r\,=\,\frac{C+\mu^2(\ln a -1)}{8 \pi G a^4}
\,=\,\frac13\,\rho_r+ \frac13\frac{|\be|{\bar F}^2}{a^4}\,.
\label{pressure}
\eeq
As one can see here, the quantum effect on the background
electromagnetic fields decreases the radiation pressure.
The dependence between the temperature and the scale factor
is given by
\beq
T\,=\,\frac{1}{a}\,\Big[\frac{45\,(C+\mu^2\ln a)}{8\pi^3G}\Big]^{1/4}
\,=\,\frac{1}{a}\,\Big[\frac{15}{\pi^2}\,
\left(\,
\rho_r^{0}\,-\,\,|\be|{\bar F^2}\ln a
\,\right)\Big]^{1/4}\,.
\label{temperature}
\eeq
Naturally, the quantum effects produce some deviation from
the usual classical formulas, namely, the $\rho_r$ is a bit
larger than $\rho_r^{0}$ for a given temperature.
\vskip 10mm

\section{Conclusions and discussions}

We have considered a cosmological applications of the
vacuum quantum effects in the radiation-dominated universe
and found that the $\be F^2$ term in the conformal anomaly
leads to a slight modification of the evolution law and the
thermal history of the universe. In the transitional period
between inflation and radiation dominated universe the
$\be F^2$ gives a non-zero contribution to the acceleration
of the universe, that is different from the classical
radiation. It would be interesting to explore further
physical consequences of this effect.

The relation for the anomaly-induced effective action in the 
radiation sector can be useful for investigating the general 
features of the gravity with anomaly-induced quantum 
corrections. In particular, it would be very interesting 
to explore the stability of the corresponding semi-classical 
solution, for a realistic particle content, that means 
non-stable Starobinsky inflation. This problem is well-known 
as a problem of stability of Minkowski and de Sitter spaces 
(see, e.g., 
\cite{stab1,stab2,stab3} and further reference therein). Our 
previous analysis also shows that the stability conditions 
for the conformal factor fo the metric
may be different for the flat space from one side and for 
the dS space from another one \cite{asta}. 
It would be very interesting 
to check out what are the conditions of stability of the 
classical solution in a general case, for different stages
of the universe expansion. The effect of radiation in the 
anomaly-induced action (ref{general solution}) is potentially 
relevant on this respect. We hope to report on this 
issue in a clase future. 

\section*{Acknowledgments}
Authors are grateful to J. Fabris and J. Sol\`a for useful
discussions at the early stage of the work. A.P. is grateful
to the Quantum Field Theory group at UFJF for the hospitality
during the short visit there. I.Sh. is grateful to CNPq,
FAPEMIG, FAPES and ICTP for partial support.

\renewcommand{\baselinestretch}{0.9}


\begin {thebibliography}{99}

\bibitem{fhh} M.V. Fischetti, J.B. Hartle, B.L. Hu,
              Phys.Rev. D20 (1979) 1757.

\bibitem{star} A.A. Starobinski, Phys.Lett. 91B (1980) 99.

\bibitem{mm}
S.G. Mamaev, V.M. Mostepanenko, Sov.Phys.-JETP 51 (1980) 9.

\bibitem{star1} A.A. Starobinski, JETP Lett.  30 (1979) 682;
JETP Lett.  34 (1981) 460.

\bibitem{ander} P.R. Anderson, Phys. Rev.  D28 (1983) 271.
Phys. Rev. D29 (1984) 615;
Phys. Rev. D32(1985) 1302;
%
Phys. Rev. D33 (1986) 1567.

\bibitem{vile} A. Vilenkin, Phys. Rev. D32 (1985) 2511.

\bibitem{anju} J.C. Fabris, A.M. Pelinson, I.L. Shapiro,
Grav. Cosmol. 6 (2000) 59;
Nucl. Phys. B597 (2001) 539.

\bibitem{hhr} S.W. Hawking, T. Hertog, H.S. Real,
Phys. Rev. D63 (2001) 083504.

\bibitem{insusy} I.L. Shapiro,
Int. J. Mod. Phys. D11 (2002) 1159, 
 arXiv: hep-ph/0103128.

\bibitem{shocom} I.L. Shapiro, J. Sol\`a,
Phys. Lett. 530B (2002) 10.

\bibitem{asta} A.M. Pelinson, I.L. Shapiro and F.I. Takakura,
Nucl. Phys. 648B (2003) 417, 
hep-ph/0208184; \ \
Nucl. Phys. B(PS) 127 (2004) 182, 
 arXiv: hep-ph/0311308.

\bibitem{asta-s}
A.M. Pelinson, I.L. Shapiro, J. Sola, F.I. Takakura,
JHEP (Proceedings Section) PRHEP-AHEP2003/033, 1-11;
 arXiv: hep-ph/0311363.

\bibitem{ana-so}
A.M. Pelinson, Int. Journ. Mod. Phys. D18  (2009) 1355.

\bibitem{fossil}
J. Sol\`a,
Journ. Phys. A41 (2008) 164066, arXiv: 0710.4151 [hep-th].

\bibitem{TUZh} E.C. Thomas, F.R. Urban, A.R. Zhitnitsky,
JHEP 0908 (2009) 043, arXiv: 0904.3779 [gr-qc].

\bibitem{book}
I.L. Buchbinder, S.D. Odintsov and I.L. Shapiro,
{\it Effective Action in Quantum Gravity}
(IOP Publishing, Bristol, 1992).

\bibitem{Poimpo} I.L. Shapiro,
{\it Effective Action of Vacuum: Semiclassical Approach},
Class. Quant. Grav. 25 (2008) 103001; arXiv: 0801.0216 [gr-qc].

\bibitem{birdav} N.D. Birell and P.C.W. Davies, {\it Quantum fields
in curved space} (Cambridge Univ. Press, Cambridge, 1982).


\bibitem{FormQED} B. Gon\c{c}alves, G. de Berredo-Peixoto
and I.L. Shapiro,
Phys. Rev. D80 (2009) 104013; \ arXiv: 0906.3837 [hep-th].

\bibitem{Mottola-08}
M. Giannotti and E. Mottola, Phys. Rev. D79 (2009) 045014, arXiv:0812.0351.

\bibitem{Coriano} R. Armillis, C. Coriano, L. Delle Rose,
Phys. Rev. D81 (2010) 085001, arXiv: 0910.3381 [hep-ph].

\bibitem{rei} R.J. Riegert, Phys.Lett. 134B (1980) 56;

E.S. Fradkin, A.A. Tseytlin, Phys.Lett. 134B (1980) 187.

\bibitem{stab1}
P.R. Anderson, C. Molina-Paris, E. Mottola, 
Phys. Rev. D67 (2003) 024026,  arXiv: gr-qc/0209075; 
Phys. Rev. D80 (2009) 084005, arXiv: 0907.0823.

\bibitem{stab2} G. Perez-Nadal, A. Roura, E. Verdaguer,
Phys. Rev. D77 (2008) 124033,  arXiv:0712.2282.

\bibitem{stab3} B.L. Hu, E. Verdaguer, 
Living Rev. Rel. 11 (2008) 3, arXiv:0802.0658.
 

\end{thebibliography}
\end{document}